\documentclass[twocolumn]{aastex62}
\usepackage{natbib}

\defcitealias{Kataria.Das.2018}{KD2018}

\graphicspath{{./}{figures/}}

\received{....}
\revised{.....}
\accepted{.....}
\submitjournal{ApJ}

\shorttitle{Bulge effect on pattern speed}
\shortauthors{Kataria $\&$ Das}

\begin{document}
\title{The Effect of Bulge Mass on Bar Pattern Speed in Disk Galaxies}

\author[0000-0002-0786-7307]{Sandeep Kumar Kataria}
\affil{Indian Institute of Astrophysics, Bangalore, \\ 560034 ,India}
\affiliation{Indian Institute of Science, Bangalore \\560012, India
}

\author{Mousumi Das}
\affiliation{Indian Institute of Astrophysics, Bangalore \\
560034, India }



\begin{abstract}

We present a study of the effect of bulge mass on the evolution of bar pattern speed in isolated disk galaxies using N-body simulations. Earlier studies have shown that disk stars at the inner resonances can transfer a significant amount of angular momentum to the dark matter halo and this results in the slow down of the bar pattern speed. In this paper we investigate how the mass of the other spheroidal component, the bulge, affects bar pattern speeds. In our galaxy models the initial bars are all rotating fast as the $\cal{R}$ parameter, which is the ratio of the corotation radius to bar radius is less than 1.4, which is typical of fast bars. However, as the galaxies evolve with time, the bar pattern speed ($\Omega_p$) slows down leading to $\cal{R} >$1.4 for all the models except for the model with the most massive bulge, in which the bar formed late and did not have time to evolve.  The rapid slowdown of $\Omega_p$ is due to the larger angular momentum transfer from the disk to the bulge, and is due to interactions between stars at the inner resonances and those in the bar. Hence we conclude that the decrease in $\Omega_p$ clearly depends on bulge mass in barred galaxies and decreases faster for galaxies with more massive bulges. We discuss the implications of our results for observations of bar pattern speeds in galaxies.  
\end{abstract}


\section{Introduction}

It is well known that both bar and bulge properties in disk galaxies change significantly from early to late type spirals along the Hubble Sequence \citep{Laurikainen_2007,Binney&Tremaine_2007}. Bars in early type spirals appear to be longer and have a more uniform intensity relative to late type spirals. Their luminosity profiles appear to be flat \citep{Diaz-Garc.2016}. This could be due to an excess of old and young stars at the bar ends, presumably due to 4:1 resonance crowding \citep{Elmegreen_1996}. Bars in late type spirals, however, have exponential profiles. Another difference is that bars in early type spirals generally extend out to corotation radii whereas bars in late type spirals are shorter and extend out to the Inner Lindlblad Resonance (ILR) radii only \citep{Elmegreen_1985}. 

The bulges in disk galaxies also show a similar variation along the Hubble Sequence. Bulges in early type spirals are more luminous and more massive than those in late type spirals and the bulge to disk luminosity ratio B/D decreases from early type to late type spirals \citep{Laurikainen_2007,Graham&Worley_2008}. The value of log(B/D) for early type galaxies (Sa-Sb) is -0.49 while for late type galaxies (Sc-Sm) it is -1.40. Thus early type spirals have long, bright bars associated with massive bulges, whereas the later type spirals have relatively shorter bars and their galaxies have smaller bulges. This correlation suggests that bar formation and evolution must be related with bulge mass \citep{Gadotti.2011}. 

In an earlier study we had examined how bulge mass affects bar formation \citep{Kataria.Das.2018} (hereafter \citetalias{Kataria.Das.2018}). We found that for a given disk scale length, bars are more difficult to form in disks with massive bulges and the bar pattern speed ($\Omega_p$) increases with bulge mass. The gravity of the central bulge can affect the bar pattern speed. This is clearly shown in Fig. 8 and Fig. 12 of KD2018. In this paper we present a more detailed study of how bulge mass affects $\Omega_p$ and especially its evolution with time ($d\Omega_{p}/dt$). There have been several theoretical studies that indicate that $\Omega_p$ slows down over time. \citet{Lynden-Bell.1979} discussed that bars capture orbits as they evolve and transfer angular momentum from the inner to the outer parts of their disks. This takes place along the spiral arms in the disk. As a result a spiral structure can produce torques which reduces the pattern speed of a bar and increases bar eccentricity. Due to this slowdown of the bar, the corotation and outer lindblad resonance (OLR) radii increases. The dynamical friction \citep{Chandrasekhar.1943} of massive dark matter halos on bars has also been shown to be an important factor in the slowdown of bar pattern speeds \citep{Sellwood.1980,Weinberg.1985,Debattista.Sellwood.1996}.  However, \citet{Weinberng.Katz.2007a} claimed that this picture of dynamical friction is not applicable to galactic potentials because of the periodic nature of orbits around the bar. This can be explained as follows. As a stellar orbit precesses around the galactic center it feels a torque due to the bar during one half of the time period and provides a torque to the bar during the other half of the time period; this results in the average torque on the star to be zero. Instead of dynamical friction, resonances between halo orbits and bar orbits may play a more important role in slowing down bar pattern speeds \citep{Weinberng.Katz.2007a}.

A massive bulge may also have a similar effect as a dark halo on $\Omega_p$ with respect to angular momentum transfer. Bars transfer a significant amount of angular momentum to their bulges \citep{Saha.etal.2012,Saha.2015}; \citepalias{Kataria.Das.2018}. In the process bulges can gain spin but the net increase depends on the mass of the bulges \citep{Sahaetal.2016}. The variation of $\Omega_p$ with galaxy type has also been studied numerically and the results indicate that bars in early type galaxies with prominent bulges have higher pattern speed \citep{Combes.Elmegreen.1993}.

The most commonly used observational technique to determine $\Omega_p$ is the Tremaine-Weinberg method \citep{Weinberg.Tremaine.1984}, which uses a widespread tracer population such as old disk stars \citep{Guo.et.al.2019}. There are other methods such as those that use the cold gas kinematics \citep{Weiner.et.al.2001,Rand.Wallin.2004,Pinol-ferrer.2014}  or the location of rings around bars \citep{Fathi.et.al.2009}. The bar morphology in photometric images at  different wavelengths can also constrain $\Omega_p$ \citep{Seiger.2018}. One of the key morphological indicators of bar pattern speed is the ratio of corotation radius to bar length  ($\cal{R}$). It is used to indicate whether a bar is fast or slow. A bar is termed fast if $\cal{R}>$1.4 and slow if otherwise. Some observational studies indicate that nearly all bars, regardless of the galaxy Hubble type, are fast bars \citep{Aguerri.et.al.2015}, whereas others suggest that $\cal{R}$ depends on the galaxy morphology \citep{RSL.2008}  as well as the dynamical age of the bar \citep{Gadotti.2011}. The observational results of $\Omega_p$ suggest that it is not just galaxy and bar morphology that plays a role in constraining $\Omega_p$, the secular evolution of bars maybe important as well. Simulations are one of the only ways in which to study this evolution. 

In this study we use N-body simulations to study the effect of bulge mass on bar pattern speeds. Our aim is to see if there is a correlation of the decrease in $\Omega_p$ with bulge mass in disk galaxies. Since a bar is a global instability its evolution can be studied better with N-body simulations
\citep{Sellwood.1980,Combes.Sanders.1981,Efstathiou.1982,Sellwood&sparke.1988, Athanassoula.Misiriotis.2002,Athanassoula.2003,Valenzuela.Klypn.2003,Machado.2012, Saha.Naab.2014,Long.etal.2014}; \citepalias{Kataria.Das.2018}. In the following sections we describe the numerical methods which are used for generating initial condition and then the disk evolution. In section \ref{Results} we present the main results from our simulations of the variation of bar $\Omega_p$ with bulge mass, the change in bar properties and the angular momentum transfer between bulge, disk and halo. In section \ref{Discussion} we discuss the main implications of our work for understanding the evolution of $\Omega_p$ in galaxies.

\section{Numerical Technique}\label{Technique}
\subsection{Initial Conditions of model galaxies} 
We have generated our initial galaxy models using GalIC \citep{Yurin.Springel.2014}. This code uses parts of the Schwartzschild method to  populate the orbits of star particles. The final distribution of disk stars approaches a target density distribution by solving the Boltzmann equation. In all of our initial models we have used $10^6$ dark matter halo particles, $10^5$ disk particles and 5x$10^4$ bulge particles.

For this study we have generated 5 models that have compact disks and concentrated bulges, hence the bar instability is triggered within a couple of Gyrs of evolution as shown in our earlier work \citepalias{Kataria.Das.2018}. The details of the models are given in Table \ref{table:Models}. The total mass of each individual galaxy is equal to 63.8 x$10^{10} M_{\sun}$. This mass corresponds to the virial velocity of the halo which is equal to 140 Kms$^{-1}$ \citep{Springel.etal.2005}. For our models, the ratio of halo, disk and bulge particle masses varies from 1:1.12:0 in Model 1 to 1:1.4:0.23 in Model 5. Also, all the bulges are initially non-rotating. The disks are locally stable as the value of the Toomre parameter Q is greater than 1. The Toomre factor varies with radius and is given by $Q(r)=\frac{\sigma(r)\kappa(r)}{3.36G\Sigma(r)}$. Here $\sigma(r)$ is the radial dispersion of disk stars, $\kappa(r)$ is the epicyclic frequency of stars and $\Sigma(r)$ is the mass surface density of the disk.

\begin{table*}
\centering
 \caption{Initial Disk Models with increasing bulge masses}
\label{tab:Model Galaxy}
 \begin{tabular}{lccccccc}
 \hline
 
Models & $\frac{M_{B}}{M_{D}}$ & $\frac{M_B}{M_T}$ & $\frac{M_D}{M_T}$&  $\frac{M_H}{M_T}$& $\frac{R_b}{R_d}$ & $Q(R_{D})$ &$M_{B}$\\
   & & & & & & &  $[10^{10} M_{\sun}]$ \\
  \hline
  Model 1 &0 & 0&  0.1&0.9 & 0 & 1.077 & 0    \\
  Model 2 & 0.05& 0.005 &0.1 & 0.895&0.169& 1.198 & 0.32  \\ 
  Model 3 &0.1 &0.01 &0.1&0.89 & 0.174&1.123&0.64 \\
  Model 4 & 0.15& 0.015 &0.1 & 0.885&0.175& 1.460 & 0.96  \\
  Model 5 & 0.2&0.02 & 0.1& 0.88& 0.180&1.171 & 1.28 \\
  \hline
  \label{table:Models}
   \end{tabular}
\begin{flushleft}
Column(1) Model name (2) Ratio of bulge to disk mass (3) Ratio of bulge to total galaxy mass (4) Ratio of disk to total galaxy mass (5) Ratio of halo to total mass (6) Ratio of half mass bulge radius to disk scale length($R_{b}/R_{d}$) (7) Toomre parameter at disk scale length $R_d$(8) Bulge mass
\end{flushleft}   

\end{table*}

The profile of the dark matter halo, which is spherically symmetric, is given as
\begin{equation}
\rho_h=\frac{M_{dm}}{2 \pi} \frac{a}{r(r+a)^3}
\end{equation}   

where a is the scale length of the halo component. This scale length is related to the concentration parameter of the NFW halo by $M_{dm}=M_{200}$ \citep{Springel.etal.2005} so that the inner shape of the halo is identical to the NFW halo. Here a and c are related as follows
\begin{equation}
a=\frac{R_{200}}{c} \sqrt{2[ln(1+c)- c/1+c]}
\end{equation}
where $M_{200}$, $R_{200}$ are the virial mass and virial radius for an NFW halo respectively. 

The density profile for the disk component has an exponential  distribution in the radial direction and a $Sech^2$ profile in the vertical direction. 

\begin{equation}
\rho_d=\frac{M_d}{4 \pi z_0 h^2} \exp\Bigg(-\frac{R}{R_d}\Bigg) Sech^2\Bigg(\frac{z}{z_0}\Bigg)
\end{equation}
where $R_d$ and $z_0$ are the radial scale length and vertical scale length respectively. 

Finally the bulge component in our models has a Hernquist density profile given by
\begin{equation}
\rho_b=\frac{M_{b}}{2 \pi} \frac{R_b}{r(r+R_b)^3}
\end{equation}   
where $M_b$, $R_b$ are the total bulge mass and bulge scale length respectively (\ref{table:Models}). Here we see that the bulges are cuspy which will allow an ILR to form at any bar pattern speed. Therefore it will prohibit swing amplification as predicted by linear theory \citep{Binney&Tremaine_2007}. It has been shown that the effect of cuspy bulges is that the ILR disappears in thick disks  \citep{polyachenko.2016}, which is the case in the present work. Apart from disk thickness there are other factors like nonlinear processes \citep{widrow.pym.dubinski2008} and the inner Q barrier \citep{Bertin2014} which can put off the effect of  an ILR.

\begin{figure*}
\centering
\includegraphics[scale=0.85]{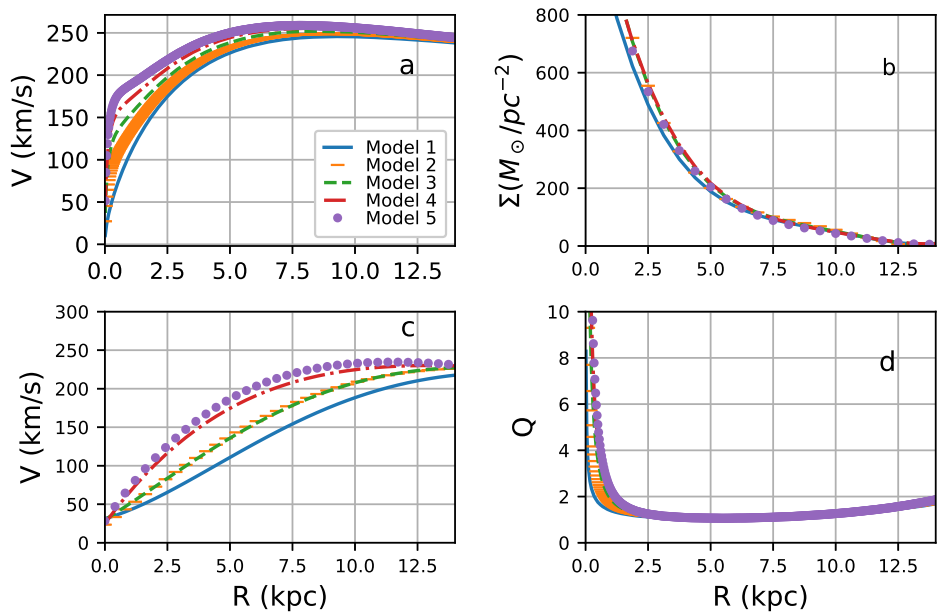}
\caption{a)~The initial rotation curves of the stellar disk; b)~the initial disk surface density; c)~the final rotation curve at 9.78~Gyr; d)~the Toomre parameter variation with radius for all the models}
\label{fig rot}
\end{figure*}

\begin{figure}
\centering
\includegraphics[scale=0.63]{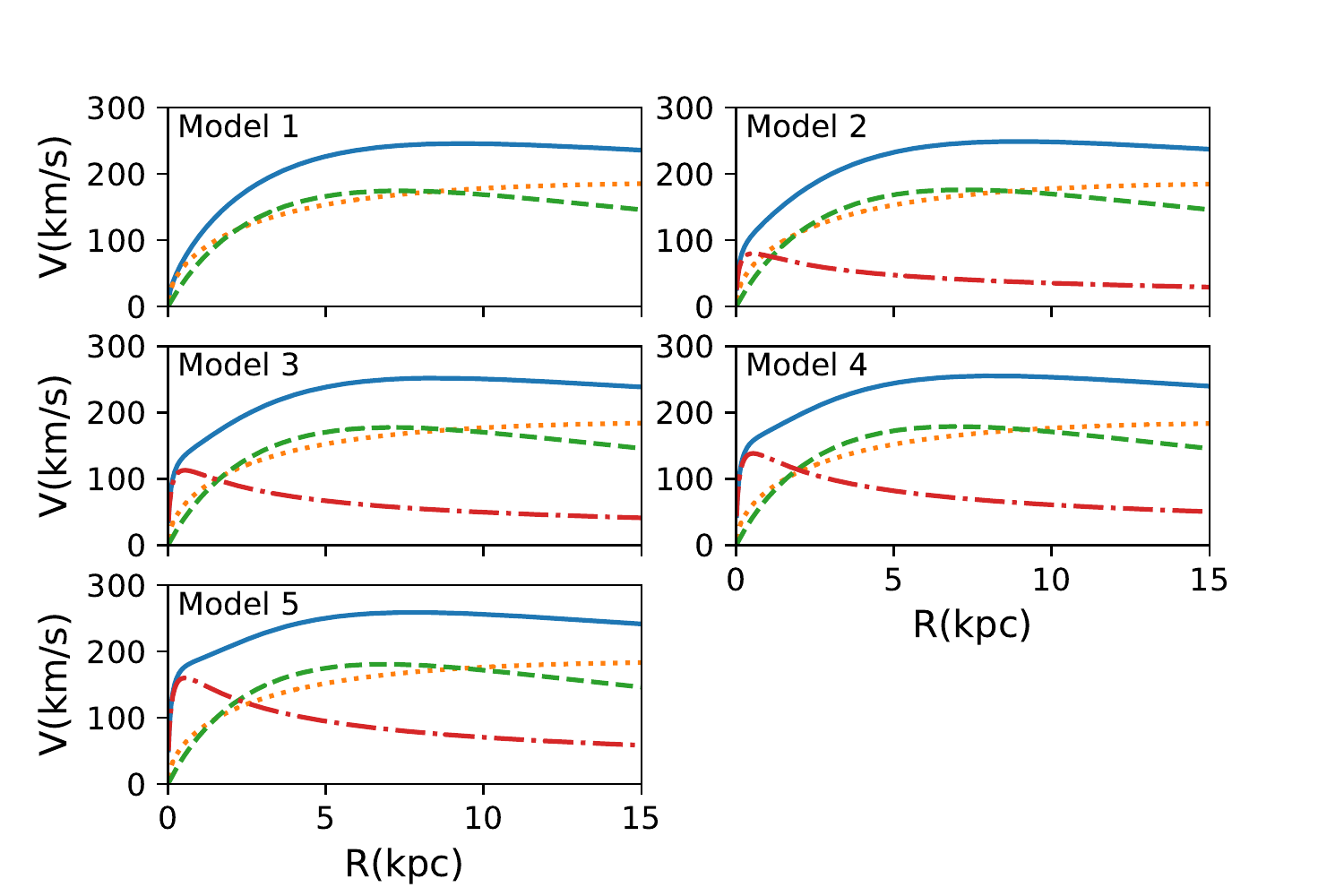}
\caption{The above plots represents the rotation curves for all of our models. The solid line is the total rotational velocity due to all the components and the rotation velocity due to the individual components is also shown. In all the plots the halo is a dotted line; the disk is a dashed line; the bulge is a dash-dot line.}
\label{fig:rotcomp}
\end{figure}

\begin{figure}
\centering
\includegraphics[scale=0.4]{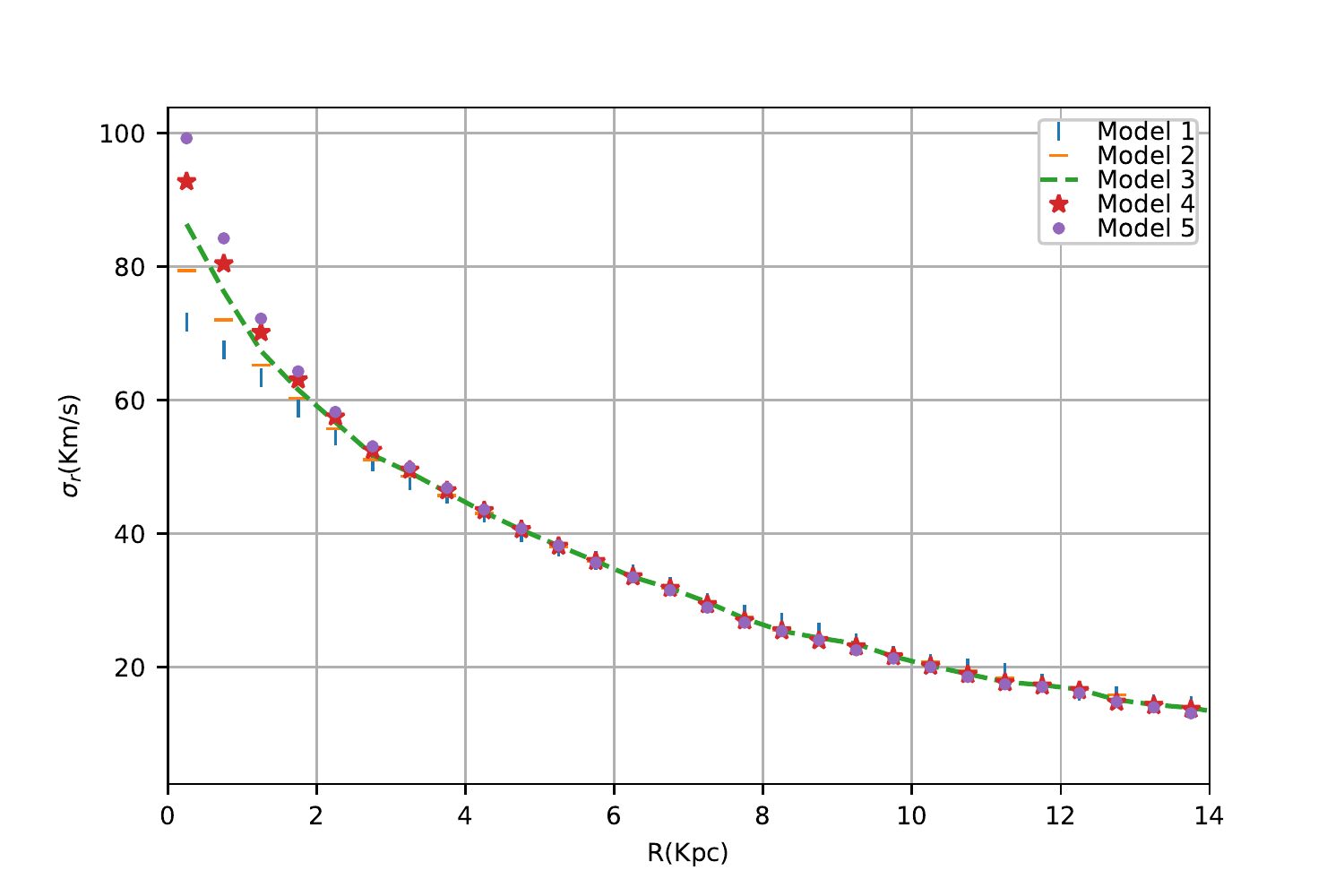}
\caption{Initial disk velocity dispersions for all the models}
\label{fig:Sigmar}
\end{figure}

Fig. \ref{fig rot} shows the initial and final rotation curves as well as the variation of initial surface density and initial Toomre parameter with radius for all of our models. We see that the inner part of the rotation curve rises with increase in bulge mass fraction which is not the case for the final evolved models which we discuss in section \ref{AME}. Fig \ref{fig:rotcomp} shows the contribution of individual components i.e. bulge, disk and halo to the total rotation curves. As expected the contribution of the bulge component to the rotation curve increases as we  keep on increasing bulge fraction. We also plot the radial velocity dispersion of the disk particle in Fig. \ref{fig:Sigmar} which shows that the inner disk velocity dispersion increases with increasing bulge mass fraction.

\subsection{Simulation Method} 

\begin{figure}
 \includegraphics[scale=0.55]{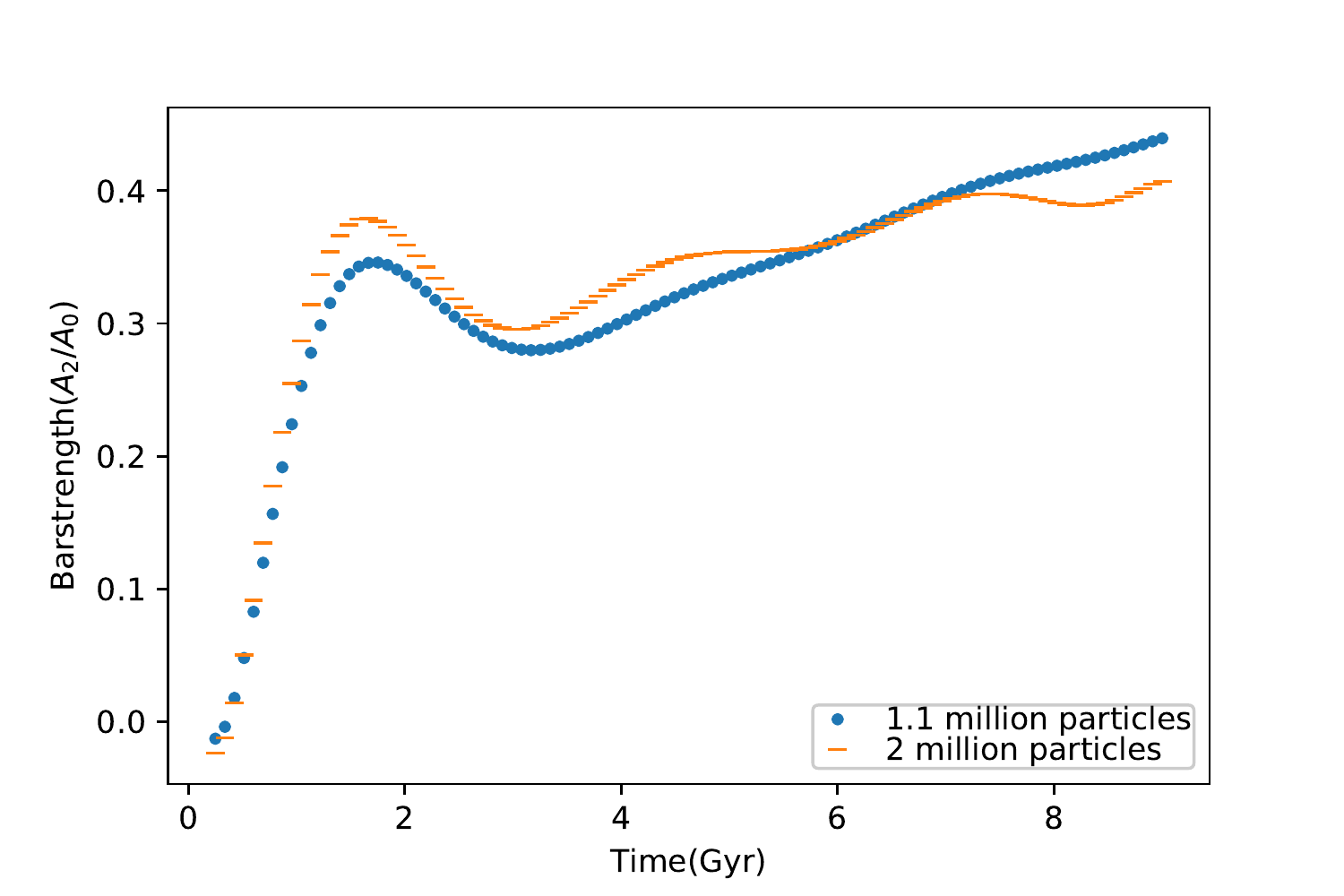}
 \caption{Bar strength evolution of Model 1 for 1.1 and 2
million particles in the simulations. A comparison of the two curves
shows that the time of growth of a bar does not depend on the number
of particles.}
 \label{fig:BSP}
\end{figure}

\begin{figure}
 \includegraphics[scale=0.55]{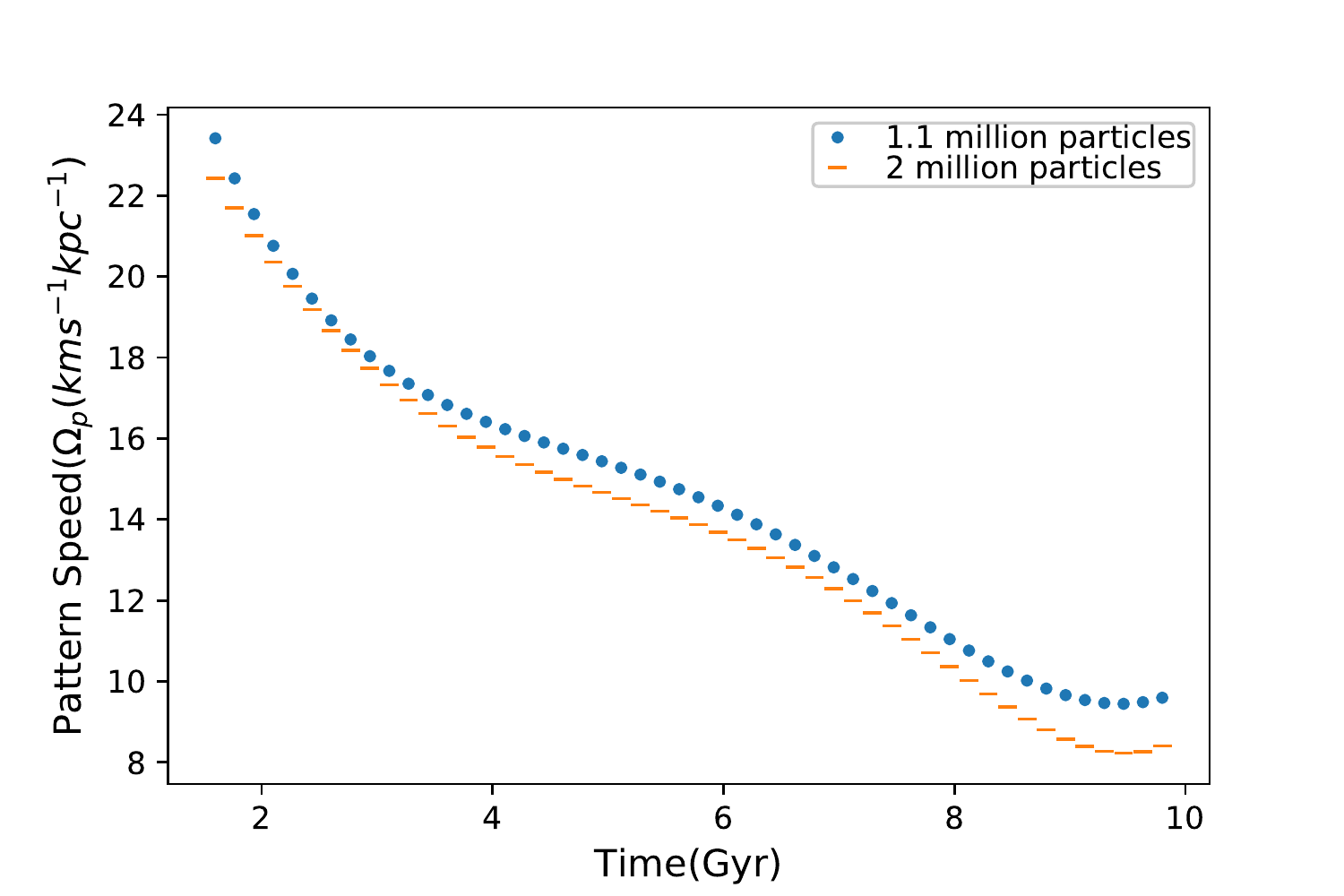}
 \caption{Pattern speed evolution of Model 1 for 1.1 and 2 million particles.}
 \label{fig:PSP}

\end{figure}

We evolved all the initial galaxy models using the GADGET-2 code \citep{Volker.2005}. This code uses the tree method \citep{Barnes.Hut.1986} to compute gravitational forces among particles. The time integration of position and velocity of particles is performed using various types of leapfrog methods. We have evolved our galaxy models up to 9.78 Gyr for conducting our pattern speed study. The opening angle for the tree is chosen as $\theta_{tot}=$0.4. The softening length for halo, disk and bulge components have been chosen as 30, 25 and 10 pc respectively. We have taken the values of the time step parameter to be $\eta< = $~0.15 and force accuracy parameter $< =$~0.0005 in most of the simulations. As a result in all of our models, the angular momentum is conserved to within 1 $\%$ of the initial value. Throughout the paper we describe our results in terms of code units. Both GalIC and Gadget-2 code have unit mass equal to $10^{10}$  $M_{\sun}$, unit distance is 1 kpc, unit velocity is 1 km/s.
We have conducted the experiment with 1.1 x $10^6$ and 2 x $10^6$ particle for the model 1 to check bar growth time scales. We find that the growth of a bar does not depend on the number of particles as shown in Figure \ref{fig:BSP}. We have also plotted the bar pattern speed in Figure \ref{fig:PSP} which also shows a similar behaviour for increasing number of particles. Therefore we finally used 1.15 x $10^6$ particles for all the simulations in this study.

\subsection{Bar strength and Pattern Speed}
Bar strength has been defined in different ways in the literature for N-body simulations  \citep{Combes.Sanders.1981,Athanassoula.2003}. In our study for defining  bar strength we have used the mass contribution of disk stars to the m=2 fourier mode.

\begin{equation}
a_2(R)=\sum_{i=1}^{N}  m_i \cos(2\theta_i)\\ \hspace{0.5cm}
b_2(R)=\sum_{i=1}^{N} m_i \sin(2 \theta_i)
\end{equation}

where $a_2$ and $b_2$ are defined in the annulus around the radius R in the disk, $m_i$ is mass of $i^{th}$ star, $\theta_i$ is the azimuthal angle. We have defined the bar strength as 
\begin{equation}
\frac{A_2}{A_0}=max \Bigg(\frac{\sqrt{a_2 ^2 +b_2 ^2}}{\sum_{i=1}^{N} m_i}\Bigg) 
\end{equation}
The above method of bar strength calculation has also been used in observations \citep{Buta.et.al_2006}, where bar strength has been calculated from the optical and near-infrared (NIR) images of disk galaxies \citep{Das.M.2008}. As in our simulations, the bar strength is derived from the variation of Fourier modes with radius in a galaxy.

We calculated the pattern speed ($\Omega_P$) of the bar by measuring the change in phase angle  $\phi=\frac{1}{2}\tan^{-1}\bigg(\frac{b_2}{a_2}\bigg)$ of the bar. This is calculated using the Fourier component in concentric annular bins throughout the disk of the galaxy. We used annular regions of width 1 kpc for disk particles only. The measured pattern speed corresponds to the bin which has the maximum value of m=2 mode i.e. bar strength.

\subsection{Angular Momentum Calculation}

The total angular momentum of the different components of a galaxy i.e. disk, bulge and halo, are measured separately. The angular momentum of a particle is calculated from the product of its mass, radial distance from galactic center and circular velocity, where the radial distance is calculated about the rotation axis of the galaxy. The time evolution of the total angular momentum of each component of the galaxy models has been discussed in section \ref{AME}.  

\subsection{Ellipticity Calculation}

We have used the {\it ellipse} task in the IRAF code to measure the bar ellipticity for all our galaxy models  \citep{Honey.etal.2016}. The initial definition of ellipticity as used by IRAF is given by

\begin{equation}
\epsilon= 1-\frac{b}{a}
\end{equation}
where b is the semi-minor axis and a is semi-major axis. In this research we have used {\it PHOTUTILS} (10.5281/zenodo.2533376), but instead of using the IRAF tasks we have used a python module of {\it Ellipse} \citep{Astropy.2018}.

Before starting this analysis we have converted the output binary files produced by the GADGET code to FITS file format \citep{Pence.2010}, which is compatible with IRAF/PHOTUTILS. To do this we generated a spatial grid in the x-y plane that has 93 bins in each direction. Then the disk particles were distributed among each of the pixels. The { \it Ellipse} task fits ellipses of increasing major axis length. We have maintained a common center for all the ellipses, which is the galaxy optical center. The position angle corresponding to each fitted ellipse remains constant until the major axis of an ellipse matches with that of the bar radius (or major axis), after which the elliptical isophotes become progressively rounder as the disk luminosity becomes more prominent. We have defined the bar ellipticity to correspond to the constant position angle within a margin of 10 degree such that there is a sharp decrease in ellipticity by $20\%$. We have defined these values as $\epsilon_{sce}$.    

\subsection{$\cal{R}$ parameter calculation}
$\cal{R}$ is the ratio of corotation radius ($R_{CR}$) to bar radius ($R_b$). 
We calculated the corotation radius by determining the distance of the lagrange point(L1/L2) from the center of the galaxy. At the lagrange point, the effective force is zero since the gravitational force balances the centripetal force in the bar rotating frame \citep{Sellwood.Debattista.2000}. We have divided the galaxy into annular bins of constant radial widths. Then for each bin we calculated the gravitational pull along the bar major axis using the potential gradient values and the centripetal force in the bar rotating frame using the $\Omega_{p}$ value and the radius of the bin. The corotation radius was then taken to be the radius at which these two forces balance each other.

\begin{figure*}
    \centering
    \includegraphics[scale=0.37]{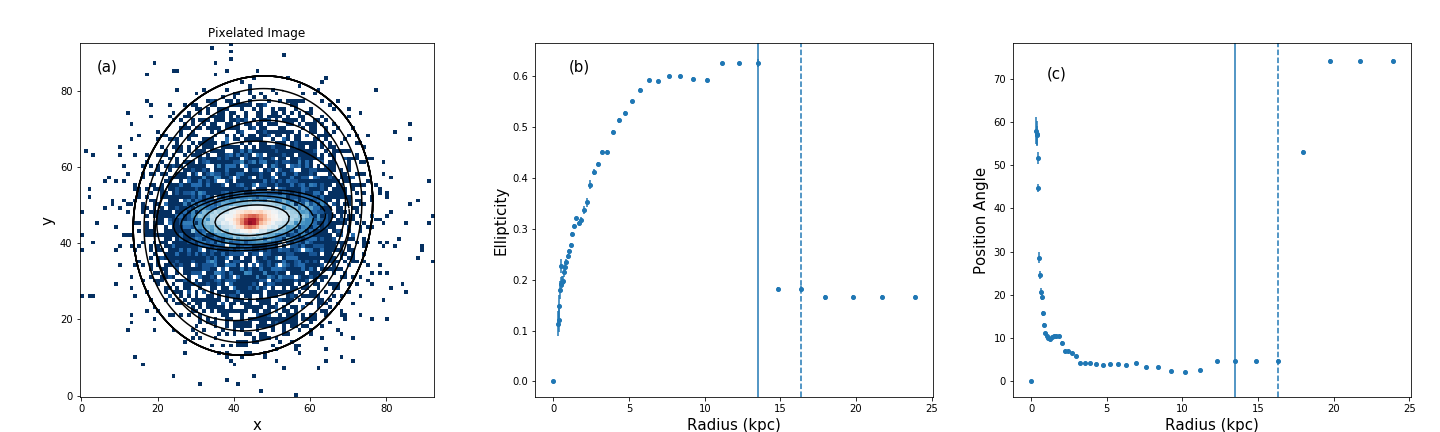}
    \caption{(a) Fitted isophotes with increasing radii Model 2 at 6 Gyr (b) Ellipticity of isophotes with increasing semi major axis of ellipses (c) Position angle (PA) of fitted ellipses with semi major axis. Here the vertical solid line represent bar length corresponding to $a_{max}$ and $a_{sce}$ definitions while dashed vertical line represent bar length corresponding to $a_{min}$ definition. }
    \label{fig:isophotes}
\end{figure*}

There are several definitions given in the literature of the measurement of bar length. These definitions correspond to the variation of the ellipticity and
position angle (PA) of the fitted isophotes, which were fitted using the {\it Ellipse} function. In the following paragraph we describe the different definitions for the semi-major bar length $a_{max}$, $a_{min}$ and
$a_{10}$.
 \\ 1. $a_{max}$ corresponds to maximum ellipticity for  $\Delta PA <10$. \citep{Erwin.2005,Marinova_Jogee_2007}  \\ 2. $a_{min}$ corresponds to minimum ellipticity for $\Delta PA <10$. \citep{Erwin.2005,Zou.Shen.2014}  \\
3. $a_{10}$ corresponds to  $\Delta PA=10$ \citep{Erwin_2003,Zou.Shen.2014}. \\ In above definitions $a_{max}$ under estimates the actual bar length while $a_{min}$ over estimates the bar length. In our study $a_{10}$ is not suitable as we are studying galaxies where the phase-angle changes sharply as we move from bar to disk isophotes or phase-on bars. We also noticed that as we fitted isophotes with increasing radii (Fig. \ref{fig:isophotes}), in a few cases the ellipses become rounder in the outer region but the PA still remains the same. This is because the position angle of circles are ill defined. Therefore we have taken the bar length to correspond to the radius where the ellipticity changes sharply by $20\%$ or more ($\epsilon_{sce}$). All the values of the final bar lengths are listed in Table \ref{table:definitions} according to all three definitions.

\begin{table*}
\centering
 \caption{Bar length and ellipticity calculation using various definitions at t=9.78 Gyr in all the models}
\label{tab:Model Galaxy}
 \begin{tabular}{l>{}c>{}c>{}c>{}c>{}c>{}c>{}c>{}c>{}c}
 \hline
 Model & $a_{min}$ & $a_{max}$ & $a_{sce}$ & $\epsilon_{min}$ & $\epsilon_{max}$ & $\epsilon_{sce}$& \\
 \hline
 
 Model 1 & 17.15& 14.19& 14.19& 0.13 & 0.748 & 0.748&  \\
 Model 2& 14.19 & 13.55 & 13.55& 0.077& 0.641 & 0.641 &   \\
 Model 3& 18.06 & 15.48 & 16.77 & 0.03 &0.637& 0.637&   \\
 Model 4& 16.125 & 16.125 & 16.125 & 0.63 &0.63& 0.63 &   \\
 Model 5& 14.19 & 10.97 & 14.19 & 0.52 &0.623& 0.6& \\
  
  \hline
  \label{table:definitions}
   \end{tabular}
\begin{flushleft}
Column(1)~Model name (2)~$a_{min}$ which corresponds to maximum ellipticity within $\Delta PA <$10 (3)~$a_{max}$ which corresponds to minimum ellipticity within $\Delta PA < $10  (4)~$a_{sce}$ which corresponds to sharp change in ellipticity by $20\%$ of previous value (5)~Ellipticity corresponding to $a_{min}$ (6)~Ellipticity corresponding to $a_{max}$ (7)~Ellipticity corresponding to $a_{sce}$. 
\end{flushleft}   

\end{table*} 

\begin{table*}
\centering
 \caption{$\Delta \Omega_p$  values for models with increasing bulge masses}
\label{tab:Model Galaxy}
 \begin{tabular}{l>{}c>{}c>{}c>{}c>{}c>{}c>{}c>{}c>{}c}
 \hline
 Model & $\Delta \Omega_p$ & initial $R_{CR}$& initial $R_b$& initial $\cal{R}$ & final $R_{CR}$& final $R_b$& final $\cal{R}$  \\
 \hline
 
 Model 1 & -16.98 & 9.275 & 8.39& 1.16 & 25.25&14.48 &1.74  \\
 Model 2&-20.94 & 7.75 & 6.45& 1.20 &24.25& 14.19& 1.71  \\
 Model 3&-24.85 & 7.4 & 5.72 & 1.29 &23.25& 15.48& 1.50   \\
 Model 4& -25.70 & 7.7 & 6.45 & 1.19 &22.75& 16.13& 1.41   \\
 Model 5& -26.11 & 6.75 & 5.71 & 1.18 &17.75& 14.19& 1.25     \\
  
  \hline
  \label{table:Domega_p}
   \end{tabular}
\begin{flushleft}
Column(1)~Model name (2)~$\Delta \Omega_p$ value (3)~initial value of the co-rotation radius $R_{CR}$ (4)~initial value of bar radius $R_b$ (5)~initial value of $\cal{R}$ where $\cal{R}=R_{CR}/R_b$  when the bar just formed (6)~final value of co-rotation radius $R_{CR}$  (7)~final value of the bar radius $R_b$ (8) ~final value of the $\cal{R}$ for models evolved up to 9.78 Gyr. 
\end{flushleft}   

\end{table*}  

\section{Results} \label{Results}

\subsection{Pattern Speed ($\Omega_p$)}
\begin{figure}
\centering
\includegraphics[scale=0.55]{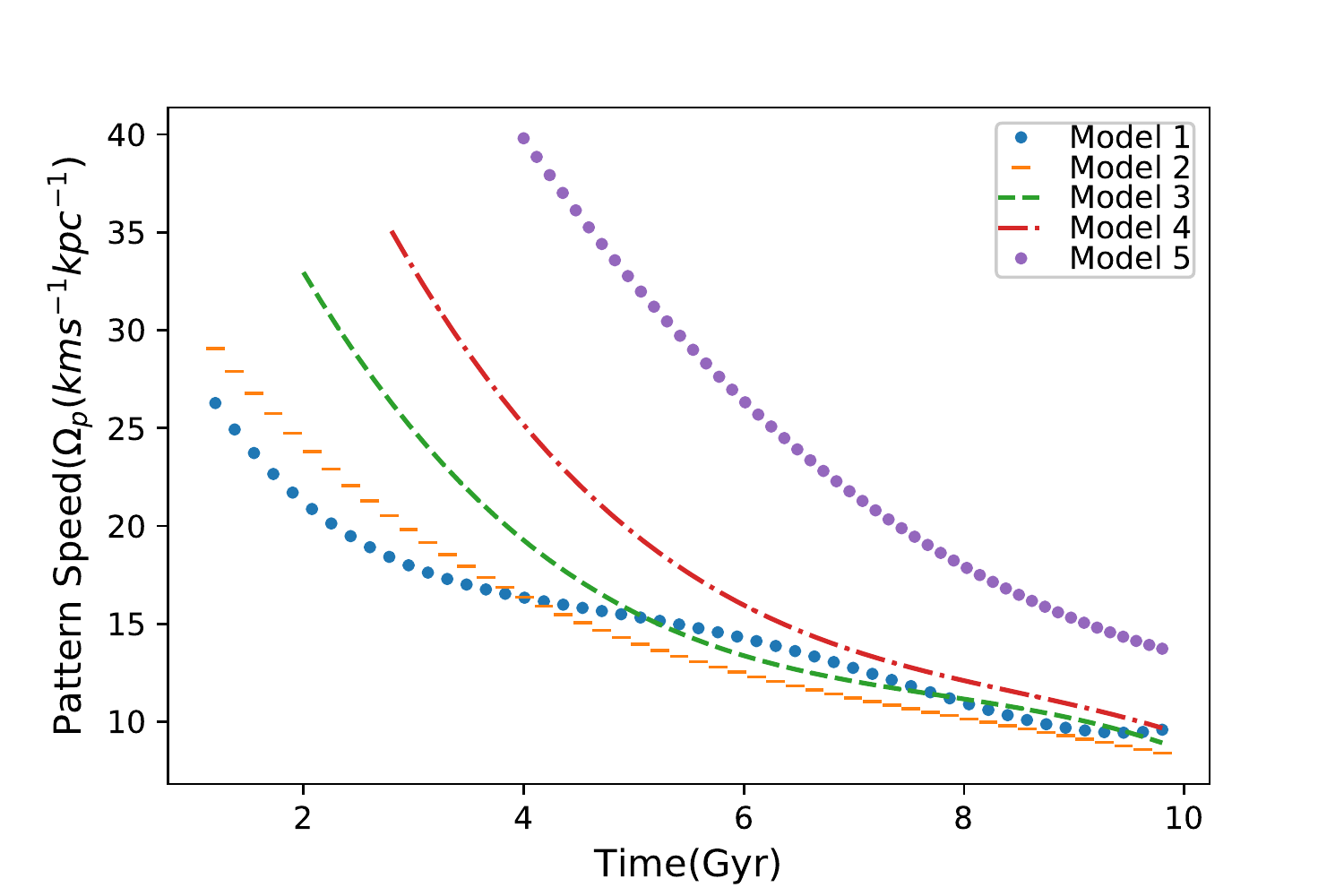}
\caption{Pattern speed evolution of all the models with time}
\label{fig:PS}
\end{figure}

\begin{figure}
\centering
\includegraphics[scale=0.55]{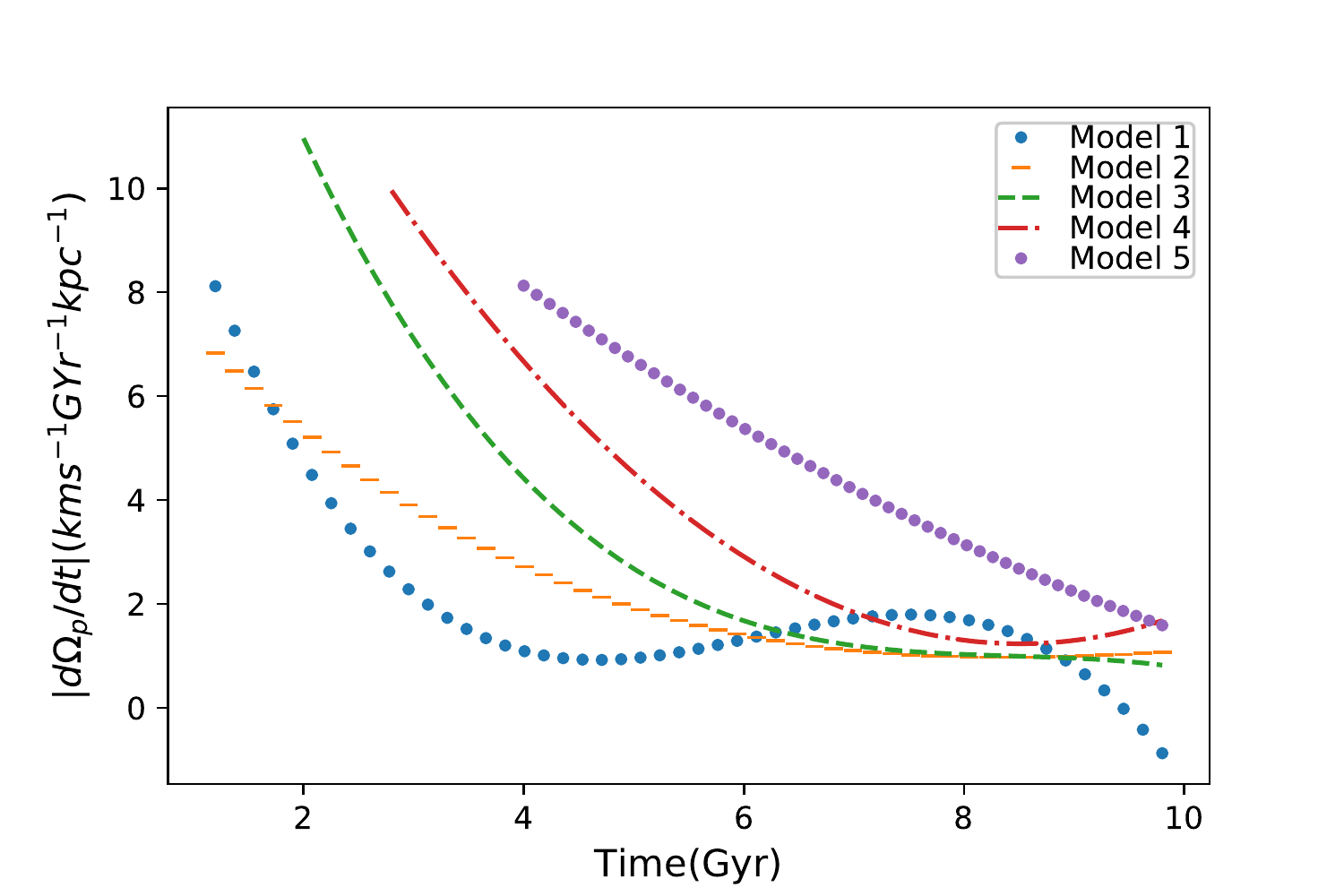}
\caption{The absolute value of Rate of change in pattern speed of bar with time for all the models}
\label{fig:PSR}
\end{figure}

The $\Omega_p$ evolution for all the models have been shown in Fig. \ref{fig:PS}. We have plotted the pattern speed only after the bar has formed. We set a cutoff on bar strength $A_2/A_0=0.2$ which is taken to be the starting point of bar formation. The bar formation timescales for Model 1, Model 2, Model 3, Model 4 and Model 5 are 1.2 Gyr, 1.2 Gyr, 2 Gyr, 2.5 Gyr and 3.96 Gyr respectively. The $d\Omega_p/dt$ has a very different form for the bulgeless galaxy (Model 1) compared to other models. The $d\Omega/dt$ falls rapidly and reaches a stable value for Model 1 but shows an increase in $d\Omega_{p}/dt$ after 6 Gyr which is not the case for other models. This correlates with faster increase in bar strength of Model 1 compared to other models that have bulges as shown in Fig. \ref{fig:BS}. Model 1 is the only disk without a bulge and so the absence of bulge does not affect its pattern speed evolution. Hence, the evolution of $d\Omega_{p}/dt$ is unlike the other models. Thus, as shown in Fig. \ref{fig:PSR} a bulge clearly has a significant effect on $\Omega_p$. 

For the other models with bulges, it is clear that the initial bar pattern speed increases with increasing bulge mass (Fig. \ref{fig:PS}). As the bars start evolving, the pattern speed decreases with time for all the models. We have plotted the rate of change in pattern speed ($d\Omega_p/dt$) in Fig. \ref{fig:PSR}. It is clear that the absolute value of the initial rate of $d\Omega_p/dt$  is larger for models with larger bulge masses as the slope of the curves becomes steeper with increasing bulge mass.

\subsection{$\Delta \Omega_p :$ Total change in pattern speed} \label{TPSC}
We have also quantified the total change in bar pattern speed since the bar formed and the parameter is given by:
\begin{equation}
\Delta \Omega_p = \int \frac{d\Omega_p}{dt} dt
\end{equation}

This quantity is an estimate of total change in bar pattern speed from the beginning stage of its formation to the final evolved state and is shown in Table \ref{table:Domega_p}. We can clearly see from Table \ref{table:Domega_p} that the absolute value of the decrease $\Delta \Omega_p$ increases with increase in bulge mass. This can be related to the higher rate of decrease in pattern speed for massive bulges as shown in Fig. \ref{fig:PSR}. Physically it can be interpreted as larger angular momentum transfer to bulge component through resonance interaction between bulge stars and the bar (see section \ref{Discussion}) and it increases with bulge mass.

\subsection{$\cal{R}$ parameter}
The $\cal{R}$ parameter is a measure of how fast a bar is rotating with respect to the disk. Fast bars correspond to $\cal{R}  <$1.4 while slow bars have $\cal{R}  >$1.4.  We see in our simulations that the models initially have $\cal{R}  <$1.4 as shown in Table~ \ref{table:Domega_p} which means that all the initial bars are fast. As we evolve our models up to 9.78 Gyr, we find that except for Model 5 which has the most massive bulge, the bars in all other models become slow and have  $\cal{R}  >$1.4. But the bar slowdown is not so prominent in Model 5 because the  bar in Model 5 has formed very late and hence the bar has not had enough time to slowdown despite facing the maximum angular transfer to the bulge, which is the most massive amongst all the models. 
Also, despite the early formation of a bar in Model 1, the $\Omega_p$ does not suffer as much slowdown as the other models (Table~\ref{table:Domega_p}). This is because model 1 does not have a bulge, hence it does not experience angular momentum transfer from the disk to the bulge through resonance interactions.

\subsection{Bar Formation Timescale and Bar Strength}
\begin{figure}
\centering
\includegraphics[scale=0.55]{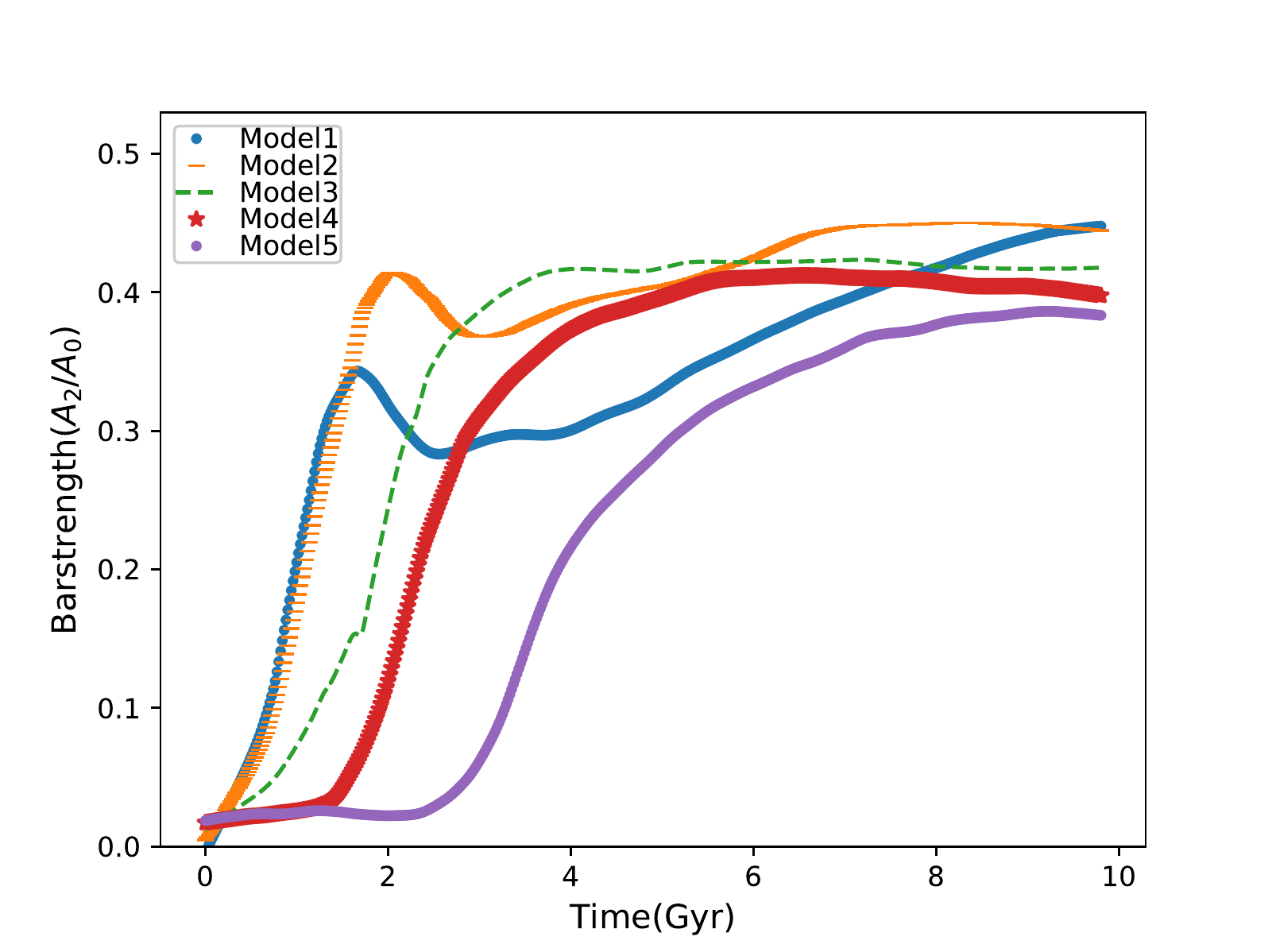}
\caption{Bar strength evolution of all the models with time}
\label{fig:BS}
\end{figure}

\begin{figure}
\centering
\includegraphics[scale=0.31]{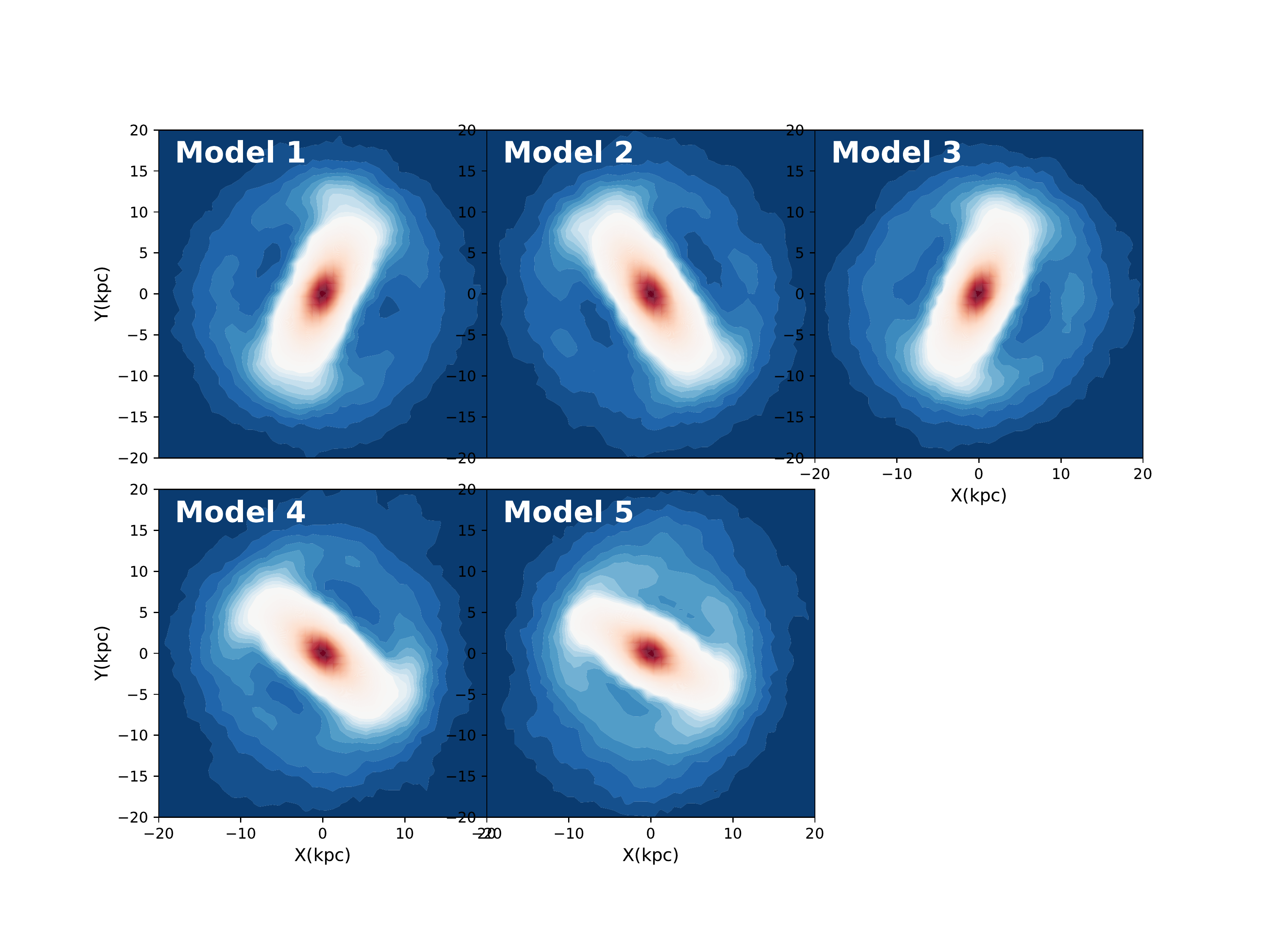}
\caption{X-Y cross section of all the evolved models at 9.78 Gyr. In this plot we have omitted bulge particle and shown only disk stars in order to focus on bar in the disk. }
\label{fig:XYCR}
\end{figure}

We have traced the evolution of bar strength with time for increasing bulge to disk mass ratios (Fig. \ref{fig:BS}). We find that the bar forms very early at  $t \sim 1.5$ Gyr for the bulgeless model (Model 1). But as the bulge mass fraction increases, the bar formation timescale increases and the bar forms later. The bar formation timescale for the less massive bulge model (Model 2) is similar to that of the bulgeless model (Model 1). The most massive bulge model (Model 5) forms a bar around $t \sim 3.96$ Gyr. This is because a more massive bulge in a galaxy center
makes a galaxy potential deeper and the disk kinematically hotter. Hence, it becomes more difficult for a bar to form. After the bar has formed, it goes into the secular evolution phase in all models. In this phase the bar strengths increase except for Model 3 and Model 4 where it is almost constant as the galaxy evolves. The final X-Y cross sections for all the models is shown in Fig. \ref{fig:XYCR} where we can see that the final bars have similar morphologies.  This trend matches that of the bar strength plot shown in Fig. \ref{fig:BS}, where nearly all the bars have similar $A_{2}/A_{0}$ values with only small variations. Thus increasing the bulge does not strongly affect the final bar strength as the final values vary by $A_{2}/A_{0} <= 0.1$

\subsection{Bar Ellipticity}
We have measured the bar ellipticity at different time intervals~: 1.98 Gyr, 3.96 Gyr, 5.94 Gyr, 7.92 Gyr and 9.78 Gyr, to study how the bar ellipticity changes with time in different galaxy models. Table \ref{table:ellipticity} shows the bar ellipticity at the above mentioned times. For the bulgeless Model 1, the ellipticity continues to increase with time. This increasing ellipticity corresponds to the continuous slowdown of $\Omega_p$ (Fig. \ref{fig:PS}). However, for the bulge dominated models (Model 2, Model 3, Model 4 and Model 5), the bar ellipticity increases continuously with time in the beginning but hovers around a mean value (Table~\ref{table:ellipticity}) as the bar evolves.
We find that the change in bar ellipticity ($\epsilon$) with time correlates with the change in bar strength (A2/A0) for all of our models. We can physically interpret this correlation in the following way; as the eccentricity of the stellar orbits in the bar increases, the bar strength also increases.
      
\begin{figure}
\centering
\includegraphics[scale=0.35]{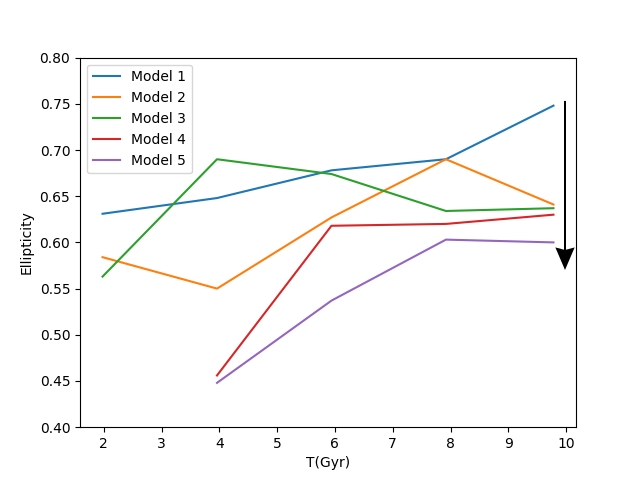}
\caption{Ellipticity evolution with time is shown for all the models. The down arrow in the figure indicates that final bar ellipticity decreases with increase in bulge mass.}

\label{fig:Ellip}
\end{figure}

This is clearly seen in Figure \ref{fig:Ellip} which shows the bar ellipticity evolution with time for all the models. We can also see that the final ellipticity of the bars at the end of the simulation decreases as bulge mass increases progressively from the bulgeless model (Model 1) to the most bulge dominated model (Model 5). This trend can be seen in bar strength values as well (Table~ \ref{table:ellipticity}). This is not surprising as the axisymmetric bulge makes the disk kinematically hotter and makes the bar orbits more circular \citep{Sellwood.1980,Athanassoula.2003,Das.et.al.2003}, causing the ellipticity to decrease. 

\begin{table*}
\centering
 \caption{Ellipticity variation with time for models with increasing bulge masses}
\label{tab:Model Galaxy}
 \begin{tabular}{lccccc}
 \hline
	 Model Name  & \hspace{1cm} t= 1.98 & \hspace{1cm}  t=3.96 &  \hspace{1cm} t=5.94 & \hspace{1cm} t=7.92 & \hspace{1cm}  t=9.78  \\
	 \hline
	 & \begin{tabular}{@{}c c@{}}$\epsilon$ &\hspace{0.25cm} $\frac{A_2}{A_0}$\end{tabular}& \begin{tabular}{@{}c c@{}}$\epsilon$ & $ \hspace{0.25cm}\frac{A_2}{A_0}$\end{tabular} & \begin{tabular}{@{}c c@{}}$\hspace{0.25cm} \epsilon$ & $\frac{A_2}{A_0}$\end{tabular}&  \begin{tabular}{@{}c c@{}}$ \hspace{0.25cm} \epsilon$ & $\frac{A_2}{A_0}$\end{tabular}& \begin{tabular}{@{}c c@{}}$ \hspace{0.25cm} \epsilon$ & $\frac{A_2}{A_0}$\end{tabular}   \\

     Model 1 & \begin{tabular}{@{}c c@{}} 0.63& 0.33\end{tabular} & \begin{tabular}{@{}c c@{}} 0.65& 0.30\end{tabular} & \begin{tabular}{@{}c c@{}} 0.68& 0.36\end{tabular} & \begin{tabular}{@{}c c@{}} 0.69& 0.42 \end{tabular} & \begin{tabular}{@{}c c@{}}0.75& 0.45\end{tabular} \\
 
     Model 2& \begin{tabular}{@{}c c@{}} 0.58& 0.43\end{tabular} & \begin{tabular}{@{}c c@{}} 0.55 & 0.38\end{tabular} & \begin{tabular}{@{}c c@{}} 0.63 & 0.42\end{tabular} & \begin{tabular}{@{}c c@{}} 0.69 & 0.45\end{tabular} & \begin{tabular}{@{}c c@{}} 0.64& 0.44\end{tabular} \\
     Model 3& \begin{tabular}{@{}c c@{}} 0.56 & 0.24\end{tabular} & \begin{tabular}{@{}c c@{}}  0.69 & 0.42 \end{tabular} & \begin{tabular}{@{}c c@{}} 0.67 & 0.42 \end{tabular} & \begin{tabular}{@{}c c@{}} 0.63 & 0.42\end{tabular} & \begin{tabular}{@{}c c@{}} 0.64& 0.42\end{tabular}  \\
     Model 4& - & \begin{tabular}{@{}c c@{}}  0.46 & 0.36\end{tabular} & \begin{tabular}{@{}c c@{}}  0.62 & 0.41\end{tabular} & \begin{tabular}{@{}c c@{}}0.62 & 0.41\end{tabular} & \begin{tabular}{@{}c c@{}} 0.63 &  0.40\end{tabular}  \\
     Model 5& - & \begin{tabular}{@{}c c@{}} 0.45 & 0.21\end{tabular} & \begin{tabular}{@{}c c@{}} 0.54 & 0.33 \end{tabular} & \begin{tabular}{@{}c c@{}}  0.60& 0.37\end{tabular} & \begin{tabular}{@{}c c@{}} 0.60 & 0.39 \end{tabular} \\
 \hline
 \label{table:ellipticity}
 \end{tabular}
  \begin{flushleft}
column(1)~Model name (2)~ellipticity and bar strength of bar at t=1.98 Gyr(3)~ellipticity and bar strength of bar at t=3.96 Gyr (4)~ellipticity and bar strength of bar at t=5.94 Gyr (5) ~ellipticity and bar strength of bar at t=7.92 Gyr (6)~ellipticity and bar strength of bar at t=9.78 Gyr 
  \end{flushleft}   

\end{table*} 

\subsection{Angular Momentum Exchange} \label{AME}

\begin{figure*}
\centering{
\includegraphics[scale=0.38]{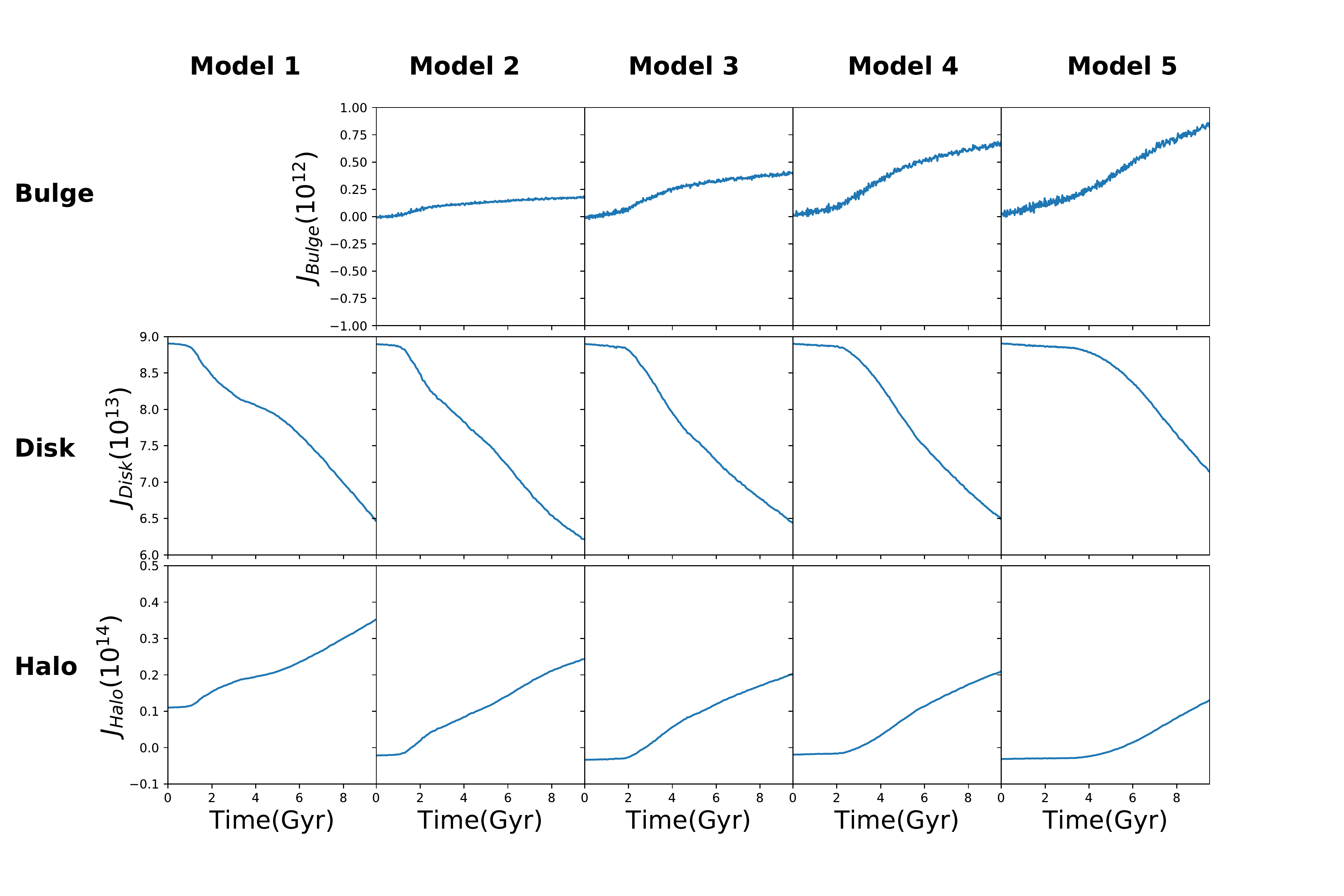}
\caption{Total Angular momentum evolution with time for individual components namely bulge, disk and halo of all the models}}
\label{fig:TAM}
\end{figure*}

During disk evolution, the angular momentum of the bulge can change with time and this may affect the bar pattern speed. To investigate this we plotted the evolution of total angular momentum with time for bulge, disk and halo components for all the models in Fig. \ref{fig:TAM}. A similar analysis was done in \citepalias{Kataria.Das.2018}. We find that the bulge and halo components always gain angular momentum with time for all the bar forming models, while the disk component always loses angular momentum The rate of change in angular momentum of the components (bulge, disk, halo) only increases after the bar has been triggered and hence the bar is important for the angular momentum exchange. As the inner disk transfers the angular momentum to the halo and bulge components, the rotation velocities of the inner disk stars decreases, this is clearly seen in the initial and final rotation curves shown in Fig. \ref{fig rot}.

We can also see that total gain in angular momentum for the bulge component increases with bulge mass as we go from Model 2 to Model 5. On the other hand, the total gain in angular momentum for the halo component decreases with increasing bulge mass. This is probably because a more massive bulge captures more disk particles and hence absorbs more angular momentum from the disk. However, it should be noted that in terms of absolute values, the halo still shows the largest increase in angular momentum. 

\section{Discussion} \label{Discussion}
In this section we discuss the implications of our results for the observations of bars. It is clear from Fig.
\ref{fig:TAM} that a bar transfers a significant amount of angular momentum to a non-rotating classical bulge and the angular momentum transfer  increases with bulge mass. The interaction is similar to the resonance between  orbits in the halo and the bar pattern speed
$\Omega_p$ \citep{Holley-Bockelmann.et.al.2005}, only in this case it is the resonant interactions between stellar orbits in the bulge 
potential and $\Omega_p$. The interaction leads to the slowdown of the bar and decline in $\Omega_p$. The dependence on bulge 
mass arises from the fact that ILR resonances are more likely to form when there is a central mass concentration such as a massive 
bulge \citep{Binney&Tremaine_2007} and the ILR concentrates the stars in a circular region. This increases the chances of resonant interactions between the bulge orbits and the bar \citep{Saha.etal.2012}, leading to the slowdown of $\Omega_p$.  

Our simulations can explain why there is no specific correlation between bulge mass and the $\cal{R}$ parameter. \cite{RSL.2008} find  that the bulge to the total disk flux (B/T) ratio is only weakly correlated to $\cal{R}$. Here  the B/T ratio  is treated as a proxy for Hubble type as suggested by previous studies \citep{Laurikainen_2007,Graham.Worley.2008}. Further, they claim that slow bars favour galaxies with low mass fraction of bulges or late type galaxies. On the other hand recent observations of 
Califa galaxies \citep{Aguerri.et.al.2015} find that most bars are fast irrespective of Hubble type.  What can be reason for different nature of the results in both the observations? 

We can explain the conflicting observations mentioned in the first paragraph using our simulation results. In our simulations, disks with more massive bulges form bars that are initially faster than those with smaller bulges. But they also tend to slowdown faster with time due to resonance interactions compared to the bars that had lower initial pattern speeds (Fig.~\ref{fig:PS} and \ref{fig:PSR}). This slowdown is more prominent in galaxies with higher bulge to disk mass fractions (see Table~\ref{table:Domega_p}). However, it is interesting to note that although the rate of slowdown is more in the case of high bulge mass fraction models, the final bar does not become as slow as $\cal{R} <$1.4. This is because the other factor which comes into play is the bar formation timescale which increases with increase in bulge mass. The bars in the models with low bulge mass fraction, despite having a lower rate of decrease in pattern speeds, have enough time to become slow bars and reach the final value of $\cal{R} >$1.4. Hence the rate of decrease in pattern speed and the timescale of bar formation play key roles in deciding whether bars are  fast or slow. Thus, our results can explain \cite{RSL.2008} if the observed bars are old (evolved for t$\approx$8 Gyr) and the Califa observations \citep{Aguerri.et.al.2015} if the observed bars are young (evolved for t$<$1Gyr).

Studies of barred galaxies suggests that bar strength decreases with increase in central velocity dispersion of stars \citep{Das.M.2008} (Table~\ref{table:ellipticity}). Since the latter increases with bulge mass and bulges are larger in early type spirals compared to late type spirals, it follows that bar strength should change along the Hubble Sequence. However, observations  suggest that there is no correlation between bar strengths and Hubble type \citep{Elmegreen.et.al.2007}. These conflicting results can be understood from the results of our simulations that  show that bar strength ($A_{2}/A_{0}$) varies for different bulge masses but the variation changes with time (Fig. \ref{fig:BS}). For example at t=4~Gyr the difference in $A_{2}/A_{0}$ between the most massive and smallest bulge is 0.2  but is 0.1 at t=10~Gyr. Thus there is a clear dependence on the secular evolution of the bars as well. So the apparently conflicting simulation and observational results can be understood as due to the fact that observations catch bars at different stages of evolution. Hence, it is not surprising that the correlations of bar strength and bulge mass that we see in our simulations is not clearly seen in observations.

\section{Summary}\label{Summary}
We have conducted an N-body study of how bar pattern speed ($\Omega_p$) in disk galaxies varies with increasing bulge mass. We summarize the main results of our work below. 

\noindent
1)~We find that the initial $\Omega_p$ of bars in galaxy disks increases with the bulge mass fraction. \\
2)~We find that initial bars in all the models are fast: $\cal{R} <$ 1.4. But as the bars evolve with time they become slower and $\cal{R} >$1.4. This trend reduces with increasing bulge mass. \\
3)~ Our simulations show that the rate of decreases in $\Omega_p$ is larger for larger bulge to disk mass fractions. The
decrease in $\Omega_p$  can be due to angular momentum transfer via resonance interactions between the bulge stars 
and the bar, and is similar to the resonant interactions between halo orbits and the bar. \\
4)~The exchange in angular momentum from disk to bulge increases with increasing bulge mass so that the most massive bulges gain the most angular momentum from the disk. This could be the reason why the bar $\Omega_p$ slows down with  increasing bulge mass. On the other hand, the exchange of angular momentum from disk to halo decreases with increasing bulge mass; however, in absolute terms, the angular momentum gain by the halo is always larger than that of the bulge\\
5)~Our simulations can explain the conflicting nature of observations of $\cal{R}$ which find both a correlation \citep{RSL.2008} and no correlation \citep{Aguerri.et.al.2015} with bulge mass or Hubble type. We can find these correlations if the observed bars are old (evolved for t$\approx$8~Gyr) and no correlation is the observed bars are young (evolved for t$<$1~Gyr). \\ 
6) The variation of bar ellipticity with time does show a correlation with bar strength. We also find that the ellipticities of the final bars show a significant decrease with increasing bulge mass.

\section{Acknowledgements}
We thank Denis Yurin and Volker Springel for providing the GalIC
code which we used for generating the initial conditions of a galaxy. We thank the authors of the Gadget-2 code (Springel 2005; Springel
et al. 2001) which we have used for this study. We thank the anonymous referee for suggestion to improve the content of this article. We are grateful to HPC facility "NOVA" at the Indian Institute of Astrophysics,
Bangalore and SERC for the CRAY facility at the Indian Institute of Science, Bangalore, where we ran our simulations.

\bibliography{sample62.bib}

\end{document}